\DocumentMetadata{
      pdfversion=2.0,pdfstandard=ua-2,
      testphase={phase-III,firstaid,math,title}
    }

\documentclass[sigconf,screen]{acmart-tagged}
\AtBeginDocument{%
  }

\usepackage{microtype} 

\usepackage{xcolor}
\usepackage[dvipsnames]{xcolor}

\setcopyright{cc}
\setcctype{by}
\acmDOI{10.1145/3757279.3785566}
\acmYear{2026}
\copyrightyear{2026}
\acmISBN{979-8-4007-2128-1/2026/03}
\acmConference[HRI '26]{Proceedings of the 21st ACM/IEEE International Conference on Human-Robot Interaction}{March 16--19, 2026}{Edinburgh, Scotland, UK}
\acmBooktitle{Proceedings of the 21st ACM/IEEE International Conference on Human-Robot Interaction (HRI '26), March 16--19, 2026, Edinburgh, Scotland, UK}
\received{2025-09-30}
\received[accepted]{2025-12-01}

\begin{document}

\title{Is Robot Labor Labor? Delivery Robots and the Politics of Work in Public Space}


%\author{ANONYMOUS AUTHOR(S)}

\author{EunJeong Cheon}
\orcid{0000-0002-0515-6675}
\affiliation{%
  \institution{Syracuse University}
  \city{Syracuse}
  \country{USA}
}
\email{echeon@syr.edu}

\author{Do Yeon Shin}
\orcid{0009-0000-9339-777X}
\affiliation{%
  \institution{University of Illinois Chicago}
  \city{Chicago}
  \country{USA}
}
\email{dshin34@uic.edu}


\begin{abstract}
As sidewalk delivery robots become increasingly integrated into urban life, this paper begins with a critical provocation: \textbf{Is robot labor labor?} More than a rhetorical question, this inquiry invites closer attention to the social and political arrangements that robot labor entails. Drawing on ethnographic fieldwork across two smart-city districts in Seoul, we examine how delivery robot labor is collectively sustained. While robotic actions are often framed as autonomous and efficient, we show that each successful delivery is in fact a distributed sociotechnical achievement—reliant on human labor, regulatory coordination, and social accommodations. We argue that delivery robots do not replace labor but reconfigure it—rendering some forms more visible (robotic performance) while obscuring others (human and institutional support). Unlike industrial robots, delivery robots operate in shared public space, engage everyday passersby, and are embedded in policy and progress narratives. In these spaces, we identify \textit{robot privilege}—humans routinely yielding to robots—and distinct perceptions between casual observers (``cute'') and everyday coexisters (``admirable''). We contribute a conceptual reframing of robot labor as a collective assemblage, empirical insights into South Korea’s smart-city automation, and a call for HRI to engage more deeply with labor and spatial politics to better theorize public-facing robots.
\end{abstract}

\ccsdesc[500]{Human-centered computing~Human computer interaction (HCI)}

\begin{comment}
\begin{CCSXML}
<ccs2012>
 <concept>
  <concept_id>00000000.0000000.0000000</concept_id>
  <concept_desc>Do Not Use This Code</concept_desc>
  <concept_significance>500</concept_significance>
 </concept>
\end{CCSXML}
\end{comment}

%\ccsdesc[500]{Do Not Use This Code~Generate the Correct Terms for Your Paper}
%\ccsdesc[300]{Do Not Use This Code~Generate the Correct Terms for Your Paper}
%\ccsdesc{Do Not Use This Code~Generate the Correct Terms for Your Paper}
%\ccsdesc[100]{Do Not Use This Code~Generate the Correct Terms for Your Paper}

\keywords{Robot labor, Delivery robots, Urban robotics, Robot privilege}
%Not, Use, This, Code, Put, the, Correct, Terms, for, Your, Pape

 % \includegraphics[width=\textwidth]{sampleteaser}
  %\caption{Seattle Mariners at Spring Training, 2010.}
  %\Description{Enjoying the baseball game from the third-base
  %seats. Ichiro Suzuki preparing to bat.}
  %\label{fig:teaser}


\maketitle

\section{Introduction}

\begin{comment}
\begin{figure}[t]
\centering
\includegraphics[width=\linewidth]{chapter/hri26-pics.png}
\caption{Delivery robots in our two field sites in Seoul. \\ Left: A robot crosses a busy intersection alongside office workers in Field A, a central business district. Right: Robots navigate a sidewalk in Field B, a smart city pilot district, passing outdoor café seating and drainage grates.}
\label{fig:field_sites}
\end{figure}
\end{comment}

\begin{figure}[t]
\centering
\includegraphics[width=\linewidth,
alt={Two field sites in Seoul. Left: A delivery robot crosses a busy intersection with office workers in a business district. Right: Delivery robots navigate a sidewalk past outdoor café seating and drainage grates in a residential area.}]{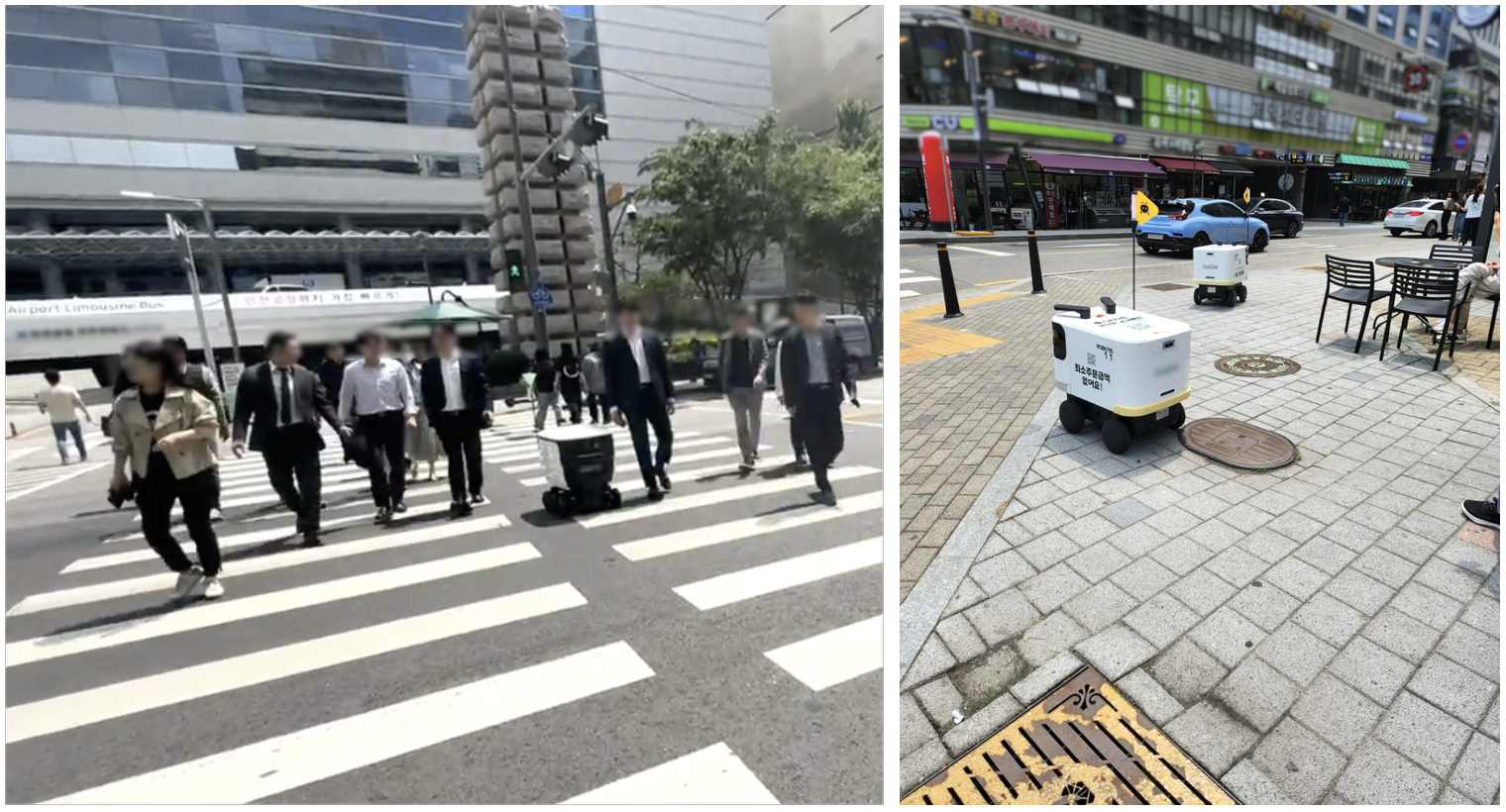}
\caption{Delivery robots in our two field sites in Seoul. Left: A robot crosses a busy intersection alongside office workers in Field A, a central business district. Right: Robots navigate a sidewalk in Field B, a smart city pilot district, passing outdoor café seating and drainage grates.}
\label{fig:field_sites}
\end{figure}

As sidewalk delivery robots become increasingly familiar presences in cities like Seoul, they prompt a deeper question: \textbf{Is robot labor labor?} This is not an inquiry about individual robotic agency or whether robots should be classified as workers. Rather, it serves as an analytical provocation: what appears as autonomous robot work is, we argue, a collective sociotechnical achievement. The question invites reflection on how robotic work is defined, legitimized, and made possible—socially, materially, and politically. While industrial machines in factories have long been framed as tools of automation—rarely discussed in terms of labor—the delivery robot blurs these boundaries. It moves autonomously through pedestrian infrastructure, communicates with passersby (e.g., ``Hello, I’m delivering, let me pass''), and completes tasks in close coordination with human and urban actors. These gestures, visibly patterned after human work, compel us to reconsider whether robotic actions are merely technical tasks (i.e., ``work'') or if they constitute a form of labor—a socially constructed, materially scaffolded activity entangled with questions of value, visibility, and control.

%%This is not just a literal inquiry about whether robots should be classified as workers. Rather, it serves as an analytical provocation, inviting 

This question holds particular significance for the HRI community, which plays a central role in designing and studying robots that increasingly participate in existing labor arrangements across public and commercial domains. As robots take on roles that support or supplant human workers, especially in public and service-oriented contexts, it becomes important to ask:  What kinds of work do robots actually perform in the world—not only functionally, but in terms of how their presence is socially recognized, valued, and legitimized? How do robots become \textit{able} to work—through whose labor, infrastructural support, and regulatory permission?  These questions gain traction in settings where delivery robots operate among multiple stakeholders—each entangled in the success or failure of robotic operation. By asking whether robot labor \textit{counts} as labor, we surface the political, symbolic, and economic functions that robots take on— what their doing \textit{means} for society, and who gets to define that meaning.

This paper draws on ethnographic fieldwork conducted across two smart-city districts in Seoul to analyze how robot labor is socially produced, sustained, and obscured. Through walk-along observation~\cite{cheon2025hello} and semi-structured interviews with a range of urban actors including shop employees, robot operators, and pedestrians, we unpack the collaborative infrastructure underlying what appears to be autonomous service delivery. We show that each “successful” delivery is not solely the result of robotic agency, but rather an emergent outcome of human labor (by shop employees), institutional coordination (through smart city initiatives and regulatory frameworks), and normative social accommodation (e.g., pedestrians yielding space). The robot’s labor, then, is not an isolated performance—but a distributed sociotechnical achievement.

We focus on delivery robots as a salient form of urban automation for three reasons. First, unlike factory or warehouse robots, they operate in open-ended, shared public environments and regularly interact with everyday publics. Second, their presence is deeply embedded in a web of political, regulatory, and economic interests—city governments, corporate operators, local businesses, and the public. Third, their social legibility—as “hardworking,” “cute,” or “commendable”—produces discourses of national progress, which obscure the unequal distribution of human labor, infrastructural strain, and affective burden sustaining the system.
This research complements and challenges prevailing framings in HRI, where robots are often studied as cooperative teammates or designable agents of trust and efficiency. By contrast, we show how robots rely on invisible human support, exploit normative asymmetries in shared space (what we term \textit{robot privilege}), and redistribute labor through ad hoc supplementation rather than substitution. Our findings suggest that automation in public life does not eliminate labor, but rather displaces, fragments, and reconfigures it.

We contribute (1) a conceptual reframing of robot labor as \textbf{a collective sociotechnical assemblage}, (2) empirical insights into how delivery robots reorder labor and infrastructure in South Korean smart-city deployments, and (3) an empirical account of how state policy and regulatory accommodations—largely underexamined in HRI—shape the conditions under which public human-robot encounters become possible. Together, these contributions call for deeper engagement with labor sociology, urban studies, and technology policy studies in the HRI research agenda. Asking whether robot labor is \textit{labor} invites us to reflect on what kinds of labor we value, who we render visible, and whose futures are being imagined through the rollout of robot labor and urban automation.

%and (3) a call to integrate labor sociology, urban studies, and critical STS into the HRI research agenda. Asking whether robot labor is labor invites us to reflect on what kinds of labor we value, who we render visible, and whose futures are being imagined through the rollout of robot labor and urban automation.
\section{Related Work}

\subsection{Delivery Robots in the Wild}

Early HRI research emphasized robots as social actors, focusing on proxemic expectations and socially appropriate behaviors in public settings \cite{jensen2005robots, takayama2009influences, yang2015experiences}. Building on this, recent studies examine how delivery robots navigate crowded sidewalks, interact with pedestrians and cyclists, and reshape public space dynamics \cite{pelikan2024encountering, duboz2025encountering, weinberg2023sharing}. These encounters generate new social norms but also frictions, as robots demand passage or cohabitate with humans in dense environments \cite{cheon2025hello}. Technical limitations—such as snow, curbs, or unpredictable street conditions—often require human intervention, revealing the labor sustaining robotic operations \cite{dobrosovestnova2022little, yu2024understanding, stedtler2025staging}. Accessibility concerns have also been raised, especially for people with disabilities, prompting calls for more inclusive and participatory robot design \cite{han2024co}. Complementing these observational studies, survey and interview-based research has explored how people perceive delivery robots. Findings show both enthusiasm for efficiency and concerns about privacy, safety, and job loss \cite{shin2024delivering, patel2024human}. Experimental work on human compliance illustrates how responses to robots’ requests depend on social context and behavioral cues, highlighting how robotic authority depends on social negotiation \cite{washburn2022exploring}. Analyses of social media and user-generated content further document how people adapt to or resist these technologies in daily life \cite{yu2024understanding}. Broader studies highlight how framing—whether robots are seen as tools or agents—shapes public attitudes, often in relation to design features such as predictability and transparency \cite{savela2018social, latikka2021attitudes, puig2023human}. Some research suggests delivery robots may be seen as fellow road users, indicating a shift toward viewing robots as part of everyday social life \cite{chi2024more}. 

Another strand of work examines the labor implications of service robots. While some debate whether robots substitute or complement human labor \cite{decker2017service}, recent studies emphasize how robots rely on invisible human work—maintenance, supervision, and social staging \cite{stedtler2025staging,dobrosovestnova2025beyond}. This co-produced autonomy raises questions about whose labor is made visible or erased. While these studies illuminate human-robot encounters in public spaces, empirical research examining public robot deployments within operational smart city contexts—and the multi-stakeholder assemblages sustaining them—remains limited. Extending these threads, we shift attention from public perceptions of robots to the collective labor that enables their operation in everyday urban life.

\subsection{Reframing Labor}
The concept and execution of labor have evolved alongside shifting social, economic, and political contexts. Historically, labor power has been assumed to originate from humans \cite{marx1981capital}. Braverman \cite{bravermanLaborMonopolyCapital1975} distinguishes human work from that of animals as “conscious and purposive,” in contrast to instinctual activity, thereby framing value-generating work as intentional and goal-directed. Noble \cite{nobleForcesProductionSocial1984} further argues that technological change reflects managerial and entrepreneurial choices rather than inherent efficiency gains. Scholars such as Braverman \cite{bravermanLaborMonopolyCapital1975} and Huws \cite{huws2003making, huws2014labor} contend that new technologies tend to displace and deskill human labor rather than liberate it, embedding the labor process in broader systems of control and alienation. Together, these perspectives highlight that labor and technology are inseparable from their social contexts and invite a rethinking of what constitutes labor and how it should be understood when assisted—or transformed—by new technologies. 

Ongoing scholarly debates have complicated this narrative by examining how robots and AI systems redistribute rather than replace human labor. Akar et al. \cite{akar2023you} and Humlum \cite{humlum2019robot} emphasize the substitutional potential of robot labor, whereas Gray and Suri \cite{gray2019ghost}, Irani \cite{irani2015difference}, and Zhang et al. \cite{zhang2025making} reveal how automated systems rely on hidden human input to function smoothly. This “ghost work” or precision labor—humans refining data, monitoring systems, and compensating for machine limitations—is often obscured from public view, reinforcing the idealized image of automation as a liberator from labor. Taylor \cite{taylorAutomationCharade2018} critiques this phenomenon with the term “fauxtomation,” describing how purportedly autonomous processes are propped up by invisible human effort. Riek and Irani \cite{riek2025future} similarly reject the idealistic narrative of automation as emancipation, underscoring instead the continued erosion and invisibility of human labor. Recently, researchers have extended this critique through the concepts of “patchwork” \cite{fox2023patchwork} and “stop-gap labor” \cite{lee2025minding}, showing how workers actively manage and repair technological breakdowns to sustain machine operations. Their findings reveal how human presence and labor power bridge the gap between the ideal of seamless automation and the reality of fragile, failure-prone systems. 

Taken together, these studies unsettle dominant assumptions about robot labor by showing how it reconfigures, rather than replaces, human work—foregrounding the need to interrogate not only what counts as labor but also whose labor remains hidden behind the promise of autonomy. Building on this body of work, our study extends these critiques to the context of public-facing, service-oriented robots by examining the labor arrangements and sociotechnical conditions that make such systems viable in practice.

\subsection{South Korea's Sociotechnical Imaginaries}

The deployment of autonomous delivery robots in Seoul is the latest chapter in South Korea's long history of state-led technological development. Robot deployment in Seoul exemplifies how South Korea's classic ``developmental state'' model  \cite{amsden1989asia, evans1995embedded} has evolved—shifting from industrial “catch-up” in semiconductors and IT to active market-creation, characteristic of an “entrepreneurial state” \cite{mazzucato2013entrepreneurial}. To understand the robot's role and public reception, it must be situated within the nation's enduring \textit{sociotechnical imaginaries}—the ``collectively held, institutionally stabilized, and publicly performed visions of desirable futures'' animated by science and technology \cite{jasanoffDreamscapesModernitySociotechnical2015}. This imaginary is performed and materialized through what has been termed “test-bed urbanism” \cite{halpern2013test}. Today, Seoul---particularly in its designated smart-city districts---functions as an urban living lab where future technologies are prototyped and showcased \cite{evansExperimentalCity2016}. This transforms the city itself into a stage for the politics of demonstration \cite{ezrahiDescentIcarusScience1990a}, where technological capability is publicly displayed to affirm state competence and construct national identity. A parallel form of this technopolitics~\cite{hechtRadianceFranceNuclear1998} is visible at Incheon International Airport, where fleets of guide and cleaning robots perform a vision of a seamless, automated “smart nation” to a global audience of travelers. The delivery robot becomes a showcase on the city street, a mobile symbol of a state-endorsed vision for a “smart city.”

Crucially, this public performance is predicated on active state intervention that prepares the ground—or ``sets the stage''—for deployment. This highlights a significant gap in the HRI literature, which has largely focused on proximate human-robot interaction and system design, while underexamining the foundational role of state policy and regulatory accommodations in making these real-world encounters possible in the first place. 

The South Korean government’s regulatory sandbox program is the key mechanism in this process. Far from simple deregulation, the sandbox is a state tool for engineering uncertainty, allowing robots to operate in a legal gray area while constructing public and institutional trust \cite{zetzsche2017regulating}. This state-led scaffolding, encompassing everything from traffic law exemptions to data privacy allowances, is part of the vast and often invisible work that underpins the robot's visible ``autonomy.'' It underscores that a robot's autonomy is a distributed accomplishment \cite{suchman2007human}, emerging from a complex assemblage of human oversight, legal exemptions, and a forgiving urban environment. This makes the robot a salient regulatory “border case,” blurring lines between property, data agent, and vehicle \cite{calo2015robotics}. The delivery robot on Seoul's streets, therefore, is better understood as a materialization of this broader national project. It represents a specific future vision that is simultaneously engineered through policy and infrastructure, performed for a public audience, and imagined within a collective national narrative.

\section{Method}

This study employed an urban ethnographic approach, drawing on the authors' combined expertise in HCI/HRI and anthropology, to investigate the real-world deployment of autonomous delivery robots. Grounded in a social constructionist epistemology and relativist ontology within the ``Big Q'' qualitative paradigm \cite{braun2013successful, braun2021one}, we adopted an inductive analytical approach to understand how robot labor is collectively produced and perceived across urban settings.

We conducted over 120 hours of intensive fieldwork across two districts of Seoul, South Korea, over a two-week period. These districts were chosen for their contrasting urban characteristics and their roles as long-running (18+ months), state-supported delivery robot testbeds.
Field A is a central business district with fast-paced pedestrian flows and heavy foot traffic (Figure~\ref{fig:field_sites}, left). It hosts Seoul’s largest urban robot deployment, part of a national initiative to demonstrate large-scale AI and 5G-based integration. The project is supported by a consortium of municipal agencies and tech corporations and places numerous robots outdoors across major intersections and commercial buildings. In contrast, Field B is a mixed-use residential and commercial area designated as a “smart city living lab” since 2019 (Figure~\ref{fig:field_sites}, right). The area functions as a city-supported testbed for various ``4th industrial revolution'' technologies, including a fleet of delivery robots operating under a regulatory sandbox exemption. More than 20 robots serve eateries, residential towers, and office buildings, traversing various micro-environments throughout the day. 

To investigate these sites, we adopted the Walk-along with Robots (WawR) methodology \cite{cheon2025hello}. Inspired by mobile methods in urban studies \cite{kusenbach2003street, vestergaard2014act, jones2021public}, WawR involves walking alongside robots to observe their real-time, unscripted encounters with the urban environment \cite{cheon2025hello}. Rather than relying solely on human informants, this method positions robots as entry points and interlocutors—entities whose movement patterns, interruptions, and engagements reveal the layered social, material, and regulatory textures of their environment. By centering the robot's operational standpoint, we observed how it navigated both physical infrastructure (e.g., curbs, grates, and obstacles) and complex social situations (e.g., crowded sidewalks, curious pedestrians, and other street workers) from a grounded, in-situ perspective. The WawR framework \cite{cheon2025hello} allowed us to link observations to contextual features such as time of day, traffic density, or local terrain. 

To complement our observations, we conducted semi-structured interviews with field operators and with owners and staff from 15 participating businesses. We also conducted over 30 informal intercept interviews with “involuntary contributors” to robot operations—those who regularly shared space with the robots—including delivery riders, sanitation workers, building managers, parcel couriers, leaflet distributors, and local residents. This study received IRB approval from Syracuse University. 

Our data corpus consists of extensive fieldnotes, photographs, videos, and transcripts. We conducted daily debriefing sessions during fieldwork to discuss initial observations and developing ideas. After transcribing all materials, we analyzed the corpus using reflexive thematic analysis \cite{braun2019reflecting, braun2021can}, beginning with organic initial coding before meeting to compare interpretations. Through iterative discussions, we collaboratively generated and refined themes around labor, spatial negotiation, and symbolic interpretations of the robots.

%We analyzed this corpus using reflexive thematic analysis, collaboratively coding for patterns in labor, spatial negotiation, and symbolic interpretations of the robots. Through iterative discussions, we clustered themes to generate the findings presented in the following sections.

\section{Findings}

%Robot’s labor is parasitic on human labor
\subsection{Human Labor behind Robot Autonomy}
Although presented as autonomous, delivery robots depend on a dispersed mesh of human labor. Our fieldwork revealed that their perceived independence relies on invisible, fragmented tasks—from street-level intervention to store-level coordination. This section highlights three key sites where such human work becomes essential: field operators, store staff, and platform-mediated oversight.

%Despite their presentation as autonomous systems, delivery robots rely heavily on a surrounding mesh of human labor. Our fieldwork revealed that the robot’s perceived independence is sustained through invisible, fragmented, and redistributed forms of work that are socially and organizationally dispersed. Across streets and storefronts, we observed a multi-layered division of labor in which the successful operation of “autonomous” robots required constant human intervention, emotional attunement, and adherence to platform-driven workflows. This section focuses on three entangled sites of this labor: field operators, store staff, and the platform interface that coordinates interactions between stores and robots.

\textit{Field Operators Performing Staged Autonomy}: Although robots appeared to navigate the streets unassisted, they were quietly accompanied by human field operators who followed them throughout their routes. These workers, typically wearing casual clothing and face masks to blend into crowds, carried smartphones and controllers to manage unforeseen situations—adjusting routes, controlling speed, and troubleshooting technical failures. One operator explained their role: \textit{“We handle operations and monitoring onsite. Developers aren’t here—just us.”}
Operators were also expected to remain unseen, deliberately making their labor invisible during customer interactions by hiding in nearby buildings and only emerging when customers struggled with pickup procedures. They performed extensive manual work disguised as automation: \textit{``[We] set them up in the morning and manually drive them to standby locations. When orders come in, the robots go out. We finish at 5 PM and return them for charging''} After deliveries, we observed operators photographing the robot’s dashboard displays to document performance metrics, noting details like the distance traveled. This human support work contradicts public narratives of robotic capability, revealing what we term \textit{``staged autonomy''}—a performance of technological self-sufficiency maintained through invisible human labor.

\textit{Transferring the Burden to Store Staff}: Our fieldwork in participating shops and restaurants shows how the presence of delivery robots reconfigured daily workflows. Staff were expected to manage new layers of digital logistics: acknowledging arrival alerts, inputting prep times, loading food into robots, troubleshooting capacity issues, and confirming delivery completion in separate robot-specific apps. One café owner remarked,\textit{ “Delivery riders just come in and grab the order, but with robots, I have to take it outside and load it myself.”} The robots’ limited cargo space frequently complicated packaging routines. Staff worried about temperature changes and food quality: \textit{“If you put hot and cold together, it just ends up lukewarm.”} Others noted the inflexibility of robots compared to human riders, who might creatively balance bulky or irregularly shaped orders: \textit{“If the bike’s box is full, human riders still manage—placing items on their laps or hanging drinks on the handlebar—but robots can’t do that.”} These logistics demanded improvisation. When orders exceeded a robot’s capacity, staff had to manually split them across multiple robots, calling in backup units via an “SOS” feature in the app. Moreover, the lack of integration between robot apps and the shop’s main POS systems introduced new manual processes and coordination burdens. \textit{“We have to check the robot app separately from our POS,”} one store owner noted. \textit{“During busy hours, it’s easy to miss something.”} 

%\textcolor{blue}{Moreover, store staff experienced a new kind of algorithmic pressure that spilled over from the platform into their physical workspace. Platforms recorded prep times, enforced delivery readiness thresholds, and triggered repeated alerts or automated calls if stores failed to respond promptly. \textit{“If I don’t open the robot lid in time,” }one worker said, \textit{“the app keeps beeping and we get a call.”} Also, the burden of coordination, once shared with human riders, now fell entirely on store staff, who had to synchronize their own rhythms with a semi-automated delivery pipeline. As one worker put it, \textit{“With riders, there’s a human understanding. With robots, the pressure is constant—and silent.”}}

\textit{Platform Oversight as Workplace Spillover}: Alongside these physical and managerial tasks, store staff experienced a new kind of algorithmic pressure that spilled over from the platform into their physical workspace. Platforms recorded prep times, enforced delivery readiness thresholds, and triggered repeated alerts or automated calls if stores failed to respond promptly. \textit{“If I don’t open the robot lid in time,” }one worker said, \textit{“the app keeps beeping and we get a call.”} These relentless notifications disrupted attention and heightened emotional stress. \textit{“It vibrates and flashes nonstop,” }another lamented. \textit{“I’m alone at the counter and can’t keep up.”} The burden of coordination, once shared with human riders, now fell entirely on store staff, who had to synchronize their own rhythms with a semi-automated delivery pipeline. As one worker put it, \textit{“With riders, there’s a human understanding. With robots, the pressure is constant—and silent.”}

\subsection{Robot Privilege in Public Space}
%Asymmetric relationship on the road and \textbf{robot privilege}
%Asymmetrical Spatial Accommodation

Across our field sites, we observed recurring asymmetries in how pedestrians responded to the presence of delivery robots. These asymmetries reorganized the micro-politics of sidewalk space and produced what we call \textbf{“robot privilege”}—a tacit, normalized expectation that humans will yield to robots, not the other way around.  
%Unlike human passersby or even dogs, robots were not perceived as requiring personal space or mutual spatial negotiation. Instead, people deferred to them as if their path were preordained.
We consistently observed that people—not robots—shoulder the burden of spatial adjustment. When a robot approaches, pedestrians typically step aside, slow down, or change direction, while the robot maintains its trajectory. Rather than initiating avoidance behaviors, the robots deploy auditory cues—a car‑like honk or a polite female voice announcing, \textit{“Hello, I’m delivering, let me pass”}—that prompt humans to clear the path. These cues function as commands that secure passage for a robot in a shared public environment. As one shop owner recalled, \textit{“I stood in front of it once and it honked at me—it wanted me to move; it had to get to its place.”} Such moments reveal how robots enact a subtle but obvious claim to space, normalizing their priority on sidewalks and establishing a form of robot privilege through everyday practice.

This asymmetric dynamic is reinforced by the robots’ selective communication. We found that horns and voice messages were directed almost exclusively at humans who blocked the robots’ paths, whereas nonhuman obstacles—dogs, signs, delivery boxes, or even other robots—were quietly detoured around without sound. For example, one robot honked repeatedly at a leaflet distributor engrossed in their smartphone, but when faced with a barking dog it made a silent detour. This design choice presumes that humans, unlike other obstacles, can and should accommodate robotic passage if properly signaled. In doing so, it places the onus for coordination squarely on pedestrians, embedding a logic of human compliance into the robots’ navigational behavior. These micro-interactions illustrate the normalization of robots’ claims to space and priority—enacted through everyday, often unspoken, expectations of human accommodation.

\subsection{Urban Rhythms, Social Perception, and Micro-Infrastructures}

Our fieldwork revealed that interactions between delivery robots and pedestrians were deeply shaped by the spatial, temporal, and infrastructural rhythms of their environments. These encounters varied significantly across neighborhoods, depending on the pace of foot traffic, time of day, and micro-level urban design—highlighting the importance of situated context in shaping perceptions, disruptions, and negotiations of robotic mobility.

In two neighborhoods within the same city, robot encounters followed markedly different patterns (Figure~\ref{fig:field_sites}). In Field A, a central business district, fast-paced pedestrians—often with their gaze downward, focused on smartphones—frequently encountered delivery robots at knee-level, leading to abrupt stops and near-collisions. Few people engaged directly with the robots, although some looked back with curiosity or filmed them post-encounter. The fast rhythm of the space left little room for interaction. As one restaurant owner noted, “\textit{People here are all office workers—during lunch, they’re in a rush and don’t want delays}.” In contrast, in Field B, a smart city pilot district—a slower, more residential area—robots elicited more intentional forms of engagement. Pedestrians included families, cyclists, and older adults, many of whom showed curiosity or empathy toward the machines. Some playfully blocked the robots or anthropomorphized their struggles. One passerby, watching a robot fail to navigate a ramp, commented, “\textit{Oh, I want to pull it out!}” Others speculated humorously about its functions—\textit{“It’s going to shoot a laser at you”}—as they touched its sensors. These localized interactions highlight the influence of neighborhood character and pace on the forms of engagement robots receive.
Temporal rhythms also mattered. The same office worker who ignored a robot during a busy noon hour might later pat it gently on the way back from lunch. Public perception was temporally situated and socially contingent, shifting with time of day, crowd density, and the affective atmosphere of the setting. Shop owners attuned to these cycles strategically blocked robot orders during peak hours: “\textit{When it gets too busy, we just stop accepting robot deliveries. It’s easier to only serve the in-store customers.}”

Beyond social rhythm, delivery robots’ navigation was deeply influenced by micro-infrastructural elements often invisible to human pedestrians. Tactile paving, slight road inclines, drainage grates, and street-level signage repeatedly disrupted their paths. In some cases, robots stalled at sidewalk ramps or were unable to cross at crosswalks due to signal cycles too short for their pace. Obstacles left by other sidewalk users posed recurring challenges. As one shopkeeper explained, “\textit{Sometimes a parked bike blocks it, and it just waits until someone moves it.}”
These observations point to the need for robot-aware urban planning, but also for adaptive robotic design that is sensitive to the frictions of everyday environments. Importantly, we observed robots occasionally diverging from human-like navigation patterns—such as taking shortcuts or aligning with pedestrian crosswalk rhythms—revealing a distinct logic of movement shaped by technical constraints. Together, these findings underscore that robotic delivery is not simply a question of route optimization or collision avoidance. Rather, it unfolds within a deeply contextual web of urban rhythms, human moods, and micro-material contingencies. For smooth integration, delivery robots must learn not only to navigate space, but also to synchronize with the social tempos and infrastructural textures of the cities they inhabit.

\subsection{The Everyday Robotics of Street Workers}

For street workers—motorcycle delivery riders, sanitation workers, parcel couriers, building managers—robots are not ``novel events'' but predictable elements within their daily work routine. Unlike casual passersby, these workers possess intimate knowledge of robots' operational details that emerges from iterative coexistence rather than formal training. A cargo delivery driver notes: ``\textit{At lunchtime, two of them come together. That's when I move my cart aside for a moment.}'' This precise understanding of time-based operational patterns demonstrates how repeated encounters generate situated knowledge. A facility manager explains: ``\textit{You know when there's a young guy following along. Then I think, `Oh, it's a test day.'}'' His ability to distinguish operational modes based on the presence or absence of human supervisors reveals the granular observational skills developed through routine exposure. These workers perform invisible articulation work \cite{strauss1985work}—temporarily securing obstacles, preemptively clearing paths, and adjusting their own movement patterns to smooth the robot's flow. (See Appendix A for extended field vignettes.)

Delivery riders do not perceive robots as immediate threats. A rider states: ``\textit{They can't operate in the rain, can they? Our work is still different.}'' His response suggests a grounded understanding of the robot's operational limits rather than a defensive stance. Another rider elaborates: ``\textit{Commercialization  [of robot delivery] seems difficult. There are too many variables. Look how small those wheels are.}'' These pragmatic assessments become evident in their everyday observations: ``\textit{Last time, it just stood there for a long time,}'' one notes, identifying unpredictable stopping as an operational problem. Another adds: ``\textit{It's too slow. People already complain that motorcycles are slow.}'' Particularly insightful is the observation that ``\textit{there are too many variables—everyone touches it at least once, poking or tapping it,}'' recognizing how public curiosity becomes an operational risk. Crucially, these workers assert: ``\textit{From a delivery perspective, we don't see it as a competitor at all.}'' A leaflet distributor captures the qualitative difference: ``\textit{That thing only goes straight. People go around.}'' For these workers, threats belong to a ``distant future,'' while present reality demands pragmatic coexistence strategies.

The linguistic choices of pedestrians versus street workers reveal different ways of seeing robots. Casual observers consistently use the term ``cute.'' Two young women passing a robot outside a coffee shop call out: ``\textit{So cute, hello!}'' while waving and making hugging gestures. Middle-aged women at an outdoor table comment: ``\textit{It's cute, it even has eyes.}'' In contrast, everyday coexisters employ the Korean term \textit{giteukhada}—loosely translated as ``admirable'' or ``clever''—a concept that combines recognition of competence with affectionate acknowledgment of effort, lacking an exact English equivalent. A liquor delivery worker states: ``\textit{It's admirable. It travels well on its own. Even when it got stuck on a curb, I wondered if it would get out by itself, and it did.}'' The distinction between ``cute'' and \textit{giteukhada} reflects more than word choice. ``Cute'' reflects aesthetic-emotional responses based on appearance, while \textit{giteukhada} contains pragmatic evaluations of function and performance. When a building security guard notes, ``\textit{Those eyes? They can't see sideways. Corners are the problem,}'' he perceives not ``cute eyes'' but sensor limitations.

\subsection{Robot as Public Spectacle and State-Sanctioned Future}

Across our intensive fieldwork in both fieldsites, we observed a consistent pattern: the robot’s presence triggered conversations about national progress and technological standing. On a Tuesday afternoon, we stood inside a convenience store in Field B, speaking with the owner, a man in his late 50s. As a robot slowly navigated the crosswalk outside, he pointed with his chin and remarked: “\textit{I’ve seen this on TV. In Shanghai, the restaurants… they have robots serving everything. Just a few human staff. That, right there, is a sign of a developed country. It’s amazing, really. It shows you how far we’ve come.}” This sentiment, which explicitly frames the robot as a benchmark of techno-national achievement, was not an isolated incident. Construction workers on their lunch break, watching a robot struggle momentarily with a curb, commented to one another, “\textit{The world has really advanced.}” This perception resonates strongly with—and reaffirms—the state’s longstanding narrative of achieving technological self-reliance and global competitiveness through innovation \cite{jasanoffDreamscapesModernitySociotechnical2015}.

The robot’s public performance is carefully staged. Both of our fieldsites are officially designated “smart city testbeds” (Field A's `Robot Street' and Field B's `Smart City Living Lab'). This designation provides the institutional scaffolding for the robot’s presence, most critically through the government’s regulatory sandbox program, which grants legal exemptions from traffic and data privacy laws that would otherwise prohibit their operation. These policy mechanisms—particularly the regulatory sandbox program—effectively transforms public sidewalks into state-legitimated stages for innovation, enabling the robots to serve both as technological trials and as public demonstrations of a state-endorsed future. 
This demonstration is deliberately constructed. Robots are consistently positioned with their front-facing “eyes” directed toward pedestrian flows—what robot operators describe as maintaining an “outward gaze.” The visual setup includes large “autonomous driving” signs and fluttering flags attached to docking stations, maximizing the scene’s visibility as a public display of technological progress. After lunch hours beneath the canopy in Field A, we often observed a routine post-operation task: robot operators photographing robots at their docking stations, capturing mileage screens and system readouts. These images serve internal reporting and potential media campaigns. Such staging creates the appearance of seamless autonomy while quietly revealing the ongoing labor that sustains the system’s operation.

Local businesses, central to this demonstration, are well aware of this setup. While some noted modest sales increases, most emphasized the promotional benefits and relational ties with the robot company. A salad shop owner in Field B put it bluntly: “\textit{Do we do it for the money? Not really. It’s for the promotional effect. People see the robot, they see our sticker on it, they get curious. It doesn’t cost us anything, so there’s no reason not to. It shows we’re part of what’s happening here.}” Other owners echoed similar views, valuing visibility—such as being featured in evening news segments—over delivery efficiency. One remarked, “\textit{Media inquiries identify us as the robot delivery location.}” These media encounters turn pilot participation into public narratives, portraying businesses as active partners in local innovation.

The relational dimension of these testbed performances becomes particularly apparent in Field B, where robot company employees have become regular patrons at participating restaurants. One café manager described the relationship as “mutually supportive,” with everyday commercial interactions  leading to eventual business collaboration. Another shop owner traced their involvement back to these personal ties: “\textit{Robot company employees have been eating here since we opened. We even partnered on employee meal vouchers. So when the robot delivery proposal came, it was hard to say no.}”

Within this moral economy of mutual recognition and institutional endorsement, the robot acts as more than a logistical device. It serves as a mobile advertisement, a symbol of public-private collaboration, and a visible commitment to smart city development. Its “labor” lies not just in moving goods, but in helping stage a desirable, technologically advanced urban future in which it is jointly supported by the state, corporations, and local merchants.

\section{Discussion}

\subsection{Rethinking Robot Labor }

%Questioning Robot Labor: Is Robot Labor Labor or Means To Mask Human Labor?
Is the delivery robot a worker? On the surface, it performs routinized tasks, navigates complex environments, and interacts with the public—all characteristics traditionally associated with labor. Yet unlike human labor, robotic actions are not exchanged for wages or embedded in contractual relations with capital. This disconnect helps explain why robotic actions, even when functionally mimicking human labor, are often not perceived as “labor” in the sociopolitical sense. Rather than settling the question of whether robot labor is labor, this section reorients the inquiry toward a more generative one: Whose labor becomes visible—or disappears—through the robot’s performance? Our findings show that delivery robots cannot perform labor autonomously but instead rely on a dense mesh of human efforts—retail staff who load orders under time pressure, field operators who troubleshoot in extreme weather, and pedestrians who yield space or redirect the robot when it gets stuck. These forms of support echo the concept of “stop-gap labor” \cite{lee2025minding} and what Taylor \cite{taylorAutomationCharade2018} terms “fauxtomation,” where the appearance of automation is sustained by invisible, often precarious, human work. 

Rather than eliminating labor, delivery robots redistribute it—often downward, invisibly, and unevenly. Our fieldwork revealed that local businesses are asked to monitor robot arrival alerts, initiate robot loading, and respond to robot alerts when delays occur. As one owner put it: “\textit{The robot’s better than a rider in some ways…but I’m the one calling when it [the robot]’s late}”—referring to delays caused by capacity limits or malfunctions. The redistribution of labor in robotic delivery extends beyond physical tasks to include managerial and emotional responsibilities. While some business owners appreciated robots for reducing interpersonal tensions with human riders, this shift reconfigures rather than eliminates emotional labor, transferring new forms of stress onto shop operators who must remain constantly attentive to robotic demands. What is marketed as “convenience” for customers and businesses often results in added responsibilities for small-scale actors, subtly pressuring them to comply with the system’s demands. The robot delivery platforms log arrival times, pick-up responsiveness, and delay rates, producing performance metrics that monitor and evaluate local retailers—echoing the algorithmic oversight found in Uber and Amazon warehouses~\cite{rosenblat2018uberland, cheon2024examining,cheon2025fulfillment}. Such arrangements deepen platform-mediated labor relations in supposedly automated systems. What is framed as convenience thus becomes a system of indirect surveillance and soft coercion, challenging the image of robotic neutrality. Far from autonomous, the robot functions as an extension of managerial control.

%According to business owners we interviewed, the robot delivery platforms log arrival times, pick-up responsiveness, and delay rates, producing performance metrics used to monitor and evaluate local retailers. If a store does not respond to an alert quickly, automated calls are triggered. These systems pressure compliance through automated monitoring, echoing the algorithmic oversight found in Uber and Amazon warehouses \cite{rosenblat2018uberland, riek2025future}. Such arrangements deepen platform-mediated labor relations in supposedly automated systems. As one café owner noted, “\textit{If we don’t load quickly enough, the alarm keeps going off—then a call comes in.}” What is framed as convenience thus becomes a system of indirect surveillance and soft coercion, challenging the image of robotic neutrality. Far from autonomous, the robot functions as an extension of managerial control.

Moreover, the robot’s deployment as a symbol of national innovation amplifies its public legitimacy while concealing operational frictions that are absorbed by human labor. During our fieldwork in Seoul’s smart city districts, we found that robots were consistently framed in signage and public narratives as signs of technological autonomy. Yet their movement remains highly fragile—as we observed, robots unable to climb curbs, ride elevators, or navigate address mismatches without human help. Despite this, public deployments were often staged as successful.  As Wright \cite{wright2023robots} and Rhee \cite{rhee2018robotic} argue, the robot functions as a signifier of progress, even as it deepens infrastructural dependence and labor displacement. These limitations, repeatedly surfaced during our field observations, not only challenge the robot’s autonomy but underscore its dependency on human intervention—complicating its symbolic role and raising questions about its actual integration into everyday urban life.

In light of these dynamics, we argue for a shift from viewing robots as bounded actors to conceptualizing them as \textbf{labor assemblages}—configurations of human and nonhuman elements that collectively sustain the appearance of robotic autonomy. As Richardson \cite{richardsonRevolutionaryRobots2015} argues, robots are never fully autonomous; they are nodes in a distributed system of social order and control. This perspective foregrounds the politics of visibility and labor valuation, showing how robot labor is embedded in broader social and infrastructural systems. As Humlum \cite{humlum2019robot}  notes, automation rarely eliminates labor; automation often produces labor market polarization and reinforces inequality by reallocating labor in more fragmented and less accountable ways. For HRI, this means moving beyond collaborative and social interaction frames of human-robot relations to engage critically with labor extraction, infrastructural dependence, and regulatory enablement. Robot labor must be understood not only by what robots do, but by the political-economic conditions that render robotic systems operational—conditions that remain deeply human.

\subsection{Robot Privilege and Reordering Public Space}
% Robot Privilege and the Challenge to Urban Norms

As delivery robots move through pedestrian environments, they do more than deliver goods—they reorganize space, elicit responses, and unsettle long-standing expectations about how public places should work. Based on our fieldwork, we introduce the concept of ``robot privilege'' to describe the emerging pattern of asymmetrical accommodation in public space: humans routinely step aside, pause, or reroute for robots, while the robots themselves rarely reciprocate. Through a combination of scripted verbal cues (``\textit{Hello, I'm delivering, let me pass by}'') and persistent forward movement, robots effectively claim right-of-way—positioning themselves as subjects to be accommodated, not as agents negotiating shared space. This pattern reflects engineering constraints but also reflects an infrastructural and normative shift with political implications. Sidewalks are already negotiated zones shared by pedestrians, cyclists, street vendors, and people with mobility constraints. The robot’s seamless passage often depends on the unspoken labor of others: those quick enough to move, aware enough to notice, or deferential enough to yield. In this sense, ``robot privilege'' is both an outcome of design and an emergent spatial politics. It reflects how engineering choices—such as one-way communication protocols or object-selective recognition—delegate the burden of adjustment to humans. Robots communicate when humans are in their path, but stay silent for dogs, delivery boxes, or other robots. This silence presumes—and enforces—a hierarchy of spatial respect.

%\textcolor{purple}{At the same time, public reactions to robots reveal an inconsistency in social treatment. Many people casually touch, poke, or open compartments on robots without hesitation—behaviors they would not direct toward human workers. Restaurant employees we interviewed also preferred handing off orders to robots rather than human couriers, citing reduced emotional fatigue. These interactions suggest that robots, despite commanding spatial deference, are not fully afforded social respect. Their ambiguous status—as non-persons that command space but lack rights—raises questions about the ethics of shared public life.}

For the HRI community, the concept of “robot privilege” foregrounds the often-unquestioned assumptions that portray public robots as neutral, efficient, and legible actors—highlighting how such framings may obscure their entanglement with existing social and spatial hierarchies. It invites us to reflect: What kinds of agency, authority, and social positioning~\cite{hou2024power} are being implicitly granted to robots by design? In turn, how do these roles reconfigure spatial hierarchies in ways that prioritize robotic movement over others’ presence? What norms are being reshaped when one actor is persistently accommodated, while others are expected to adjust? Rather than treating robot behavior as neutral or purely functional, we---HRI researchers and designers---must attend to how such systems redistribute agency and reinforce asymmetries in shared environments~\cite{cheon2022robots,cheon2022working}. Doing so requires attention not only to design but also to policy, which we take up in Section 5.5.

%Policy frameworks must also evolve. Current regulations tend to focus narrowly on robot safety and operational reliability. But equitable deployment demands more: ensuring that robots do not displace or inconvenience already-marginalized publics, that they are subject to accountability standards, and that their presence serves broader civic interests—not just corporate logistics. As robots become embedded in everyday public life, robot privilege offers a critical lens for rethinking spatial justice, design responsibility, and the politics of public infrastructure.

\subsection{Reframing Cuteness through Everyday Robot Labor}

As illustrated in Section 4.4, passing pedestrians predominantly described delivery robots as ``cute.'' Yet among those who coexisted with robots in their everyday workspaces—delivery couriers, sanitation workers, building managers, shopkeepers, and loading-dock staff—a different affective register emerged. These everyday coexisters consistently used the Korean term \textit{giteukhada} (admirable) to describe robots that reliably completed their tasks. This difference in word choice reveals a situated understanding of the robot’s actions as meaningful work rather than performative charm. It highlights how perceptions of robots are shaped by people’s varying exposure, proximity, and involvement in their everyday functioning. 

HRI research has long treated “cuteness” as a core design asset, linking favorable first impressions to features such as baby-schema proportions, rounded bodies, and friendly gestures~\cite{guidi2025you,nakayama2016perception,song2021effect,chen2023research}. The broader concept of \textit{kawaii}~\cite{shiomi2024feeling,shiomi2025differential,okada2022kawaii} has extended this conversation, especially in East Asian contexts, to include fondness and tenderness~\cite{wang2024kawaii,nittono2022psychology}. While some studies highlight cuteness as a tool for engagement (e.g.,~\cite{bjorling2018teen,caudwell2019ir}) or note its potential drawbacks (e.g.,~\cite{voysey2025prevalence,caudwell2019home}), they rely on initial-encounter experiments that emphasize physical traits and immediate perceptual responses~\cite{chen2023research,shiomi2023two,lv2021does,kimura2025two}—failing to capture how perceptions shift through sustained coexistence and how contextual knowledge shapes what people see in robots.

In contrast, street workers' use of \textit{“giteukhada”} centers on the robot’s perceived ability to complete tasks reliably and navigate the unpredictable conditions of the city. Rather than stemming from physical design or movement aesthetics, this recognition of the robot's competence aligns with a sociological understanding of labor—as the ability to perform socially meaningful, value-generating tasks. The expression also subtly encodes a hierarchical relationship: in Korean, it is often used by seniors to praise the diligence of juniors, such as a child completing an errand. Thus, the compliment is not only affective but evaluative and situated—marking the robot as a subordinate helper whose contributions are noticed and assessed.

Importantly, this differentiated perception underscores the uneven distribution of knowledge in robot deployments. Those who live or work alongside robots accumulate layered, practical understandings that casual observers cannot access. The distinction between “cute” and \textit{``giteukhada''} is thus not merely linguistic but positional—emerging from one’s proximity to and dependence on the robot’s functioning. Street-level workers act as informal maintainers, troubleshooters, and co-coordinators of robotic systems, yet their insights remain excluded from mainstream  HRI design and policy frameworks.
Recognizing this disparity calls for a reorientation in HRI. Rather than privileging user experience defined by customers or first-time encounters, designers and policymakers must consider the lived expertise of those who sustain the robots’ everyday functionality. Situated knowledge from enduring coexistence—not momentary impressions—should shape the future of urban robotics.

\subsection{Staging Robot Labor in Public Deployment}
%Delivery robots do more than navigate sidewalks—they mobilize state agendas, reframe national imaginaries, and reorganize urban labor. While Section 5.2 examined how delivery robots reshape spatial hierarchies and public norms, here we shift focus to the broader labor regimes and symbolic infrastructures that underlie these machines. The question of whether robot labor is “labor” prompts us to reconsider not only how robots function, but also how their labor is co-constructed and dependent on layered, often invisible human and institutional efforts. As our fieldwork revealed, the robot’s seamless movement relies on a distributed network of actors: shop employees who load meals under time pressure, operations staff who reposition and monitor the robots, and city officials who enable testing through regulatory sandboxes. As Richardson \cite{richardsonRevolutionaryRobots2015} argues, robots are never fully autonomous—they are nodes in a distributed system of social order and control. %In this light, the question \textit{“Is robot labor labor?”} becomes less about the robot itself, and more about the collective labor that disappears behind its seamless operation.

The symbolic power of delivery robots compounds the invisibility of supporting labor. %The symbolic power of delivery robots further compounds this invisibility. 
In Seoul’s smart city testbeds, the robot functions as an emblem of national competence and urban innovation. Residents and business owners spoke of them as proof that \textit{“this is what a developed country looks like.”}---projecting broader discourses of national development and technological modernity \cite{jasanoffDreamscapesModernitySociotechnical2015}. %This is automation not only as labor substitution, but as nation branding. 
%These expressions mark the robot as a floating signifier—projecting broader discourses of national development and technological modernity \cite{jasanoffDreamscapesModernitySociotechnical2015}. 
But these narratives rest on performances of what we termed ``staged autonomy.'' We observed operations staff photographing robots, positioning them to face outward, and surrounding them with `autonomous driving' signage and flags. As Shapin and Schaffer \cite{Shapin1985-SHALAT-3} remind us, technical success is often \textit{staged} as much as achieved. These performances align the state's regulatory sandbox, corporate deployment strategy, and on-the-ground logistics to render robot deployment publicly legible and politically expedient—while displacing critical questions about labor redistribution, who controls the infrastructures enabling automation, and who is accountable when things go wrong.

%This has critical implications for HRI. The field often evaluates robots in terms of task success, cooperation, or ease of use. However, our findings reveal that such metrics overlook the broader technopolitical roles robots play in urban life---as symbols of national progress, embedded in state regulation, and nodes in commercial logistics. For example, local businesses did not adopt robots solely for efficiency. Their engagement was shaped by concerns about visibility, relational obligations, and reputational positioning—echoing Thompson’s notion of moral economy~\cite{thompson1971moral}. We argue that HRI must shift toward treating robots as technopolitical artifacts whose deployment redistributes responsibility, visibility, and value. This requires drawing more actively from labor sociology, urban studies, and policy analysis to examine: who benefits from automation’s promises, who bears its burdens, and how regulatory frameworks shape these distributions. Robot labor is not only about what machines do, but how we collectively decide whose work counts, whose does not, and what visions of progress we choose to uphold.

%—criteria that reflect a view of robots as bounded technical systems. 

\subsection{Design and Policy Implications}

Our findings point to several directions for HRI design and policy. First, the concept of ``robot privilege'' calls for a shift from yield-based to negotiation-based navigation. Rather than designing robots that demand right-of-way through auditory commands, designers might explore movement patterns and multimodal cues that invite reciprocal adjustment—positioning robots as participants in shared space rather than privileged actors claiming passage. 

Second, we propose stakeholder-differentiated interfaces that reveal rather than conceal the labor assemblages sustaining robotic systems. Operators could access intervention histories showing patterns of human support, while public-facing displays might indicate operational modes (e.g., ``remote assistance active''), managing expectations about autonomy and acknowledging the distributed labor involved. 

Finally, policy frameworks must evolve. Policymakers might reconsider whose interests robotic deployment currently serves. Existing regulations focus narrowly on safety and operational reliability, yet our findings suggest that equitable deployment demands more—for instance, sandbox programs could incorporate labor impact assessments for participating businesses who absorb new coordination burdens, and deployment guidelines could require spatial impact reviews that evaluate effects on already-marginalized sidewalk users. Such transparency efforts could be further supported by regulatory requirements—for example, mandating that operators publicly report intervention rates and workforce conditions—ensuring that robotic presence serves broader civic interests, not just corporate logistics. As robots become embedded in everyday public life, ``robot privilege'' offers a critical lens for rethinking spatial justice, design responsibility, and the politics of public infrastructure.

%Finally, policy frameworks must also evolve. Current regulations tend to focus narrowly on robot safety and operational reliability. But equitable deployment demands more: ensuring that robots do not displace or inconvenience already-marginalized publics, that they are subject to accountability standards, and that their presence serves broader civic interests—not just corporate logistics. As robots become embedded in everyday public life, robot privilege offers a critical lens for rethinking spatial justice, design responsibility, and the politics of public infrastructure.
\section{Conclusion}

This paper asked: Is robot labor labor? We showed that robot labor is best understood as a collective sociotechnical achievement sustained by human effort, institutional support, and social accommodation. This has critical implications for HRI. The field often evaluates robots in terms of task success, cooperation, or ease of use. However, our findings reveal that such metrics overlook the broader technopolitical roles robots play in urban life---as symbols of national progress, embedded in state regulation, and nodes in commercial logistics. We argue that HRI must shift toward treating robots as technopolitical artifacts whose deployment redistributes responsibility, visibility, and value. This requires drawing more actively from labor sociology, urban studies, and policy analysis to examine: who benefits from automation’s promises, who bears its burdens, and how regulatory frameworks shape these distributions. Robot labor is not only about what machines do, but how we collectively decide whose work counts, whose does not, and what visions of progress we choose to uphold.

%—criteria that reflect a view of robots as bounded technical systems. 

%For example, local businesses did not adopt robots solely for efficiency. Their engagement was shaped by concerns about visibility, relational obligations, and reputational positioning—echoing Thompson’s notion of moral economy~\cite{thompson1971moral}. 


%\begin{acks}
%To Robert, for the bagels and explaining CMYK and color spaces.
%\end{acks}

\balance
\bibliographystyle{ACM-Reference-Format}
\bibliography{delivery-robot}

%\clearpage

%\appendix
%\input{chapter/appendix}

%%

\end{document}